\documentclass[draft,grl]{AGUTeX}

\usepackage{lineno}\linenumbers*[1]
\modulolinenumbers[5]

 \usepackage[dvips]{graphicx}
 \setkeys{Gin}{draft=false}
\authorrunninghead{CASELLA ET AL}

\titlerunninghead{Productive Ligurian Pool}

\authoraddr{E. Casella, P. Tepisch
CIMA Research Foundation, Via Armando Magliotto, 2 - 17100 Savona, Italy(elisa.casella@cimafoundation.org)}

\authoraddr{X. Couvelard, CCM-Center for Mathematical Sciences, University of Madeira, Campus da Penteada, 9000-390, Funchal, Madeira, Portugal (xavierc@uma.pt)}

\authoraddr{R. M.A. Caldeira, CIIMAR-Interdisciplinary Centre of Marine and Environmental Research, Rua dos Bragas, 289, 4050 - 123 Porto, Portugal (rcaldeira@ciimar.up.pt)}

\begin{document}

%
%

\title{The Productive Ligurian Pool}
%
%

%
%



 \authors{E. Casella, P. Tepisch, X. Couvelard, R. M. A. Caldeira, \altaffilmark{1}}


\altaffiltext{1}{Corresponding author}

%
%


\begin{abstract}

In contrast with the behavior of the eddies in the open-ocean, the sub-mesoscale eddies generated in the constricted Ligurian Basin (NW Mediterranean), are unproductive but their combined effect, arranged in a Ôrim-like fashionÕ, contributes to the containment of a Productive Ligurian Pool (PLP). Data derived from MODIS satellite sensor showed persistent higher chlorophyll concentrations in the centre of the basin, concurrent with high EKE values in its surroundings, derived from AVISO altimetry merged products. This suggested that this 'productive pool' is maintained by the intense (sub)mesoscale eddy activity in the rim. Numerical realistic experiments, using a Regional Ocean Model System, forced by MERCATOR and by a high-resolution COSMO-l7 atmospheric model, also showed that most of the sub-mesoscale eddies, during 2009 and 2010, are concentrated in the rim surrounding the basin, contributing to the formation of a basin-scale cyclonic gyre. We hypothesized that the interaction between eddies in the rim might contribute to import of nutrients into the pool in two ways: (i) by advection of nutrients from the nearby coastal regions into the pool; (ii) by concentrating eddy upwelled nutrients inside the pool; or by a combination thereof. 

\end{abstract}

%
%

%

\begin{article}

%
%

\section{Introduction}
\begin{figure}
\center
\includegraphics[scale=0.70]{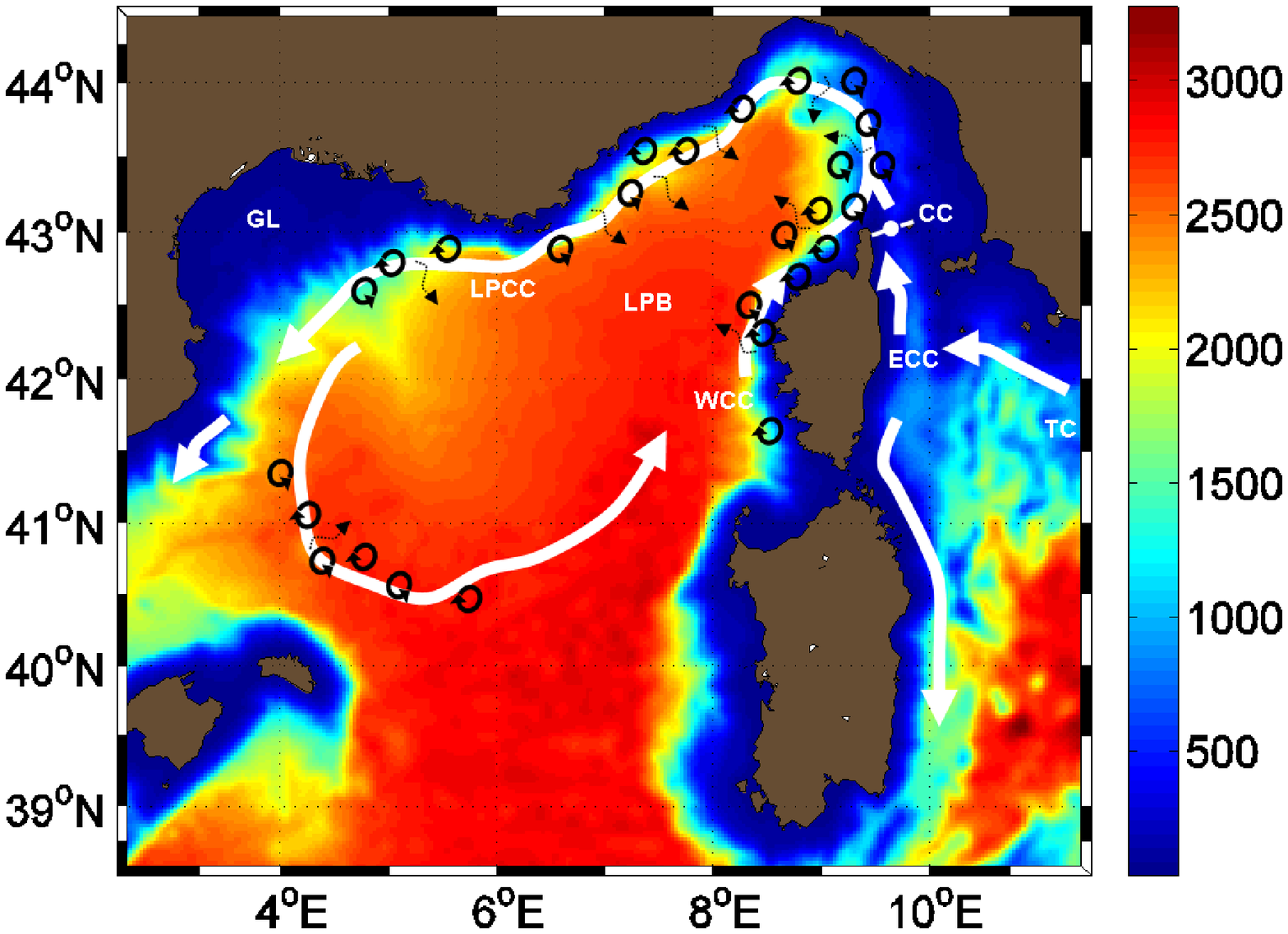}
\caption{Bathymetry (meters) and general circulation of the  Ligurian-Proven\c{c}al Basin. TC - Thyrenian Current; ECC - East Corsica Current; WCC - West Corsica Current; CC - Corsica Channel; LPB - Ligurian Proven\c{c}al Basin; LPCC - Ligurian Proven\c{c}al Catalan Current; GL - Golf of Lion. The circular patterns represent the rim-eddies and the small arrows the nutrient entrainment transport process.}
\label{map}
\end{figure}

The northwest part of the Mediterranean Sea, the Ligurian Sea, is known as an oligotrophic region \citep[e.g.][]{Nezlin_etal_2004_JGR, OrtenzioAcal_2009_Bio}, and therefore any supply of nutrients is extremely valuable to sustain the whole trophic chain. The biological production of the Ligurian Sea is dominated by a marked seasonal cycle \citep{Arone_1994} and a significant inter-annual variability \citep{LaViolette_1994, Marty_etal_2002_DSR}. Nevertheless, the system heterogeneity is closely associated with local and regional hydrodynamical factors, especially those responsible for a high variability of the mixed layer depth and for the transport processes \citep{Nezlin_etal_2004_JGR,Casella_etal_2011_JMS}. The regional hydrodynamics is dominated by a cyclonic circulation system feed by the East Corsica Current (ECC) and the West Corsica Current (WCC) which is known as the Liguro-Proven\c{c}al-Catalan (LPC) current (Figure \ref{map}). The LPC-current flows along the costal slopes of Italy, France and Spain, and is affected by instability processes which generate (sub)mesoscale eddies, capable of inducing relatively intense shelf-edge flows, producing significant dynamical heterogeneity \citep{Millot_1991_DAO}.The presence of mesoscale eddies in the Western Mediterranean Sea is well documented \citep{Santoleri_etal_1983, Marullo_Salusti_1985_DSR, Gasparini_1999_JMS, RobinsonLeslie_2001, Echevin_etal_2003_AG, Casella_etal_2011_JMS}. Nevertheless, the peculiar feature of the Northwest Mediterranean Sea emphasized herein is that, in contrast to many other marginal seas rich in mesoscale eddy activity, the nearshore zones are poor in surface chlorophyll \citep{Nezlin_etal_2004_JGR}. The LPC-current and the presence of mesoscale eddy activity, is reflected in the high values of EKE (Figure \ref{chleke}). In this discussion, we focus on  role of a rim of eddies along the Ligurian Pronven\c{c}al basin, for the maintenance of a cyclonic gyre, and for the containment of the Productive Ligurian Pool (PLP, hereafter).

\section{Data and methods}

\subsection{Satellite products}

Eddy Kinetic Energy (EKE) was computed from altimetry data. Surface velocities were computed from weekly merged products of absolute dynamic topography (ADT), at $1/8^{\circ}$ resolution on a Mercator projection, distributed by AVISO (www.aviso.oceanobs.com). ADT maps are obtained by merging measurements from all available altimeter missions \citep{Ducet_etal_2000_JGR}. Combining data from different missions significantly improves the estimation of mesoscale signals \citep{Pascual_etal_2006_GRL}.We used the processed series which considers up to 4 satellites at a given time, thus it has the best possible regional sampling. Furthermore, the period of our study (2009-2010) has one of the best altimetric coverage with a four-satellite constellation (Jason 1, Jason 2, Envisat and Cryostat). ADT is obtained adding along track Sea Level Anomaly to the Mean dynamic Topography \citep{RioHernandez_2004_JGR}. 

Chlorophyl data was derived from the MODIS-Moderate Resolution Imaging Spectroradiometer flying onboard of the Aqua platform. MODIS data was processed by the Ocean Biology Processing Group (OBPG) at Goddard Space Flight Center. Merged Level-3 chlorophyll products are being created routinely for daily, 8-day, monthly, seasonal and annual time periods, processed at a 9km spatial resolution. Although seasonal and monthly averages were analyzed, only the annual average product is shown, in order to emphasize the persistence of the PLP event during 2009 and 2010.

\subsection{The Ligurian-Proven\c{c}al Regional Ocean Modeling System}
The Regional Ocean Modeling System (ROMS) \citep{Shchepetkin_2003_JGR, ShchepetkinMcWilliams_2005_OM} was configured for the Northwestern part of the Mediterranean Sea. ROMS solves the primitive equations based on the Boussinesq approximation. In the Ligurian model solutions for 2009 and 2010, the model domain extends from $37.8^{\circ}$ N to $44.5^{\circ}$ N and from $2^{\circ}$ E to $16.5^{\circ}$ E. The bottom topography is derived from a 30 arc-second resolution database GEBCO08. In the model version adopted here, there are two open lateral boundaries: at $37.8^{\circ}$ N and $2^{\circ}$ E. We have chosen a horizontal resolution of $1/32^{\circ}$. At this resolution, the Rossby radius of deformation (of the order of $5 -12 km$ in the whole Mediterranean and for different seasons, see \citet{Grilli_Pinardi_1998} is resolved and consequently the model configuration is adequate to simulate mesoscale structures. The model grid has 35 vertical levels with vertical refinement near the surface, to obtain a satisfactory representation of the surface layer and the euphotic zone. At the open lateral boundaries, the model is forced with temperature, salinity and velocity fields obtained from the MERCATOR product PSY2V3 (www.mercator-ocean.fr). MERCATOR, has an horizontal spatial resolution of $1/12^{\circ}$, with daily outputs.

At the sea surface, the regional ocean circulation model was forced with the monthly mean climatologies of heat and freshwater fluxes, derived from the Comprehensive Ocean-Atmosphere Data Set, COADS\citep{daSilva_etal_1994}. For the atmospheric momentum, wind-stress was extracted from the Limited Area Model Italy (COSMO-I7). COSMO-I7 \citep{Montani_etal_2003_NLPG}, is a non-hydrostatic and fully compressible numerical weather prediction model, which is a regional version of the Lokal Model \citep{Schattler_Doms_2000}, regularly used for operational and research applications. The COSMO-I7, 3 hourly solutions, have an horizontal resolution of $1/16^{\circ}$. Validation of the COSMO-I7 wind fields has been widely performed \citep{Steppeler_etal_2003_MAP}. As shown in \citet{Casella_etal_2011_JMS}, forcing the Ligurian ocean circulation model with high-resolution winds, substantially contributed to the formation of mesoscale and sub-mesoscale eddies, which is an important characteristic of the regional dynamics \citep{Gasparini_1999_JMS, RobinsonLeslie_2001, Echevin_etal_2003_AG, Nezlin_etal_2004_JGR}.

\subsection{ROMS data comparison}
Non-spite the scarce availability of public data, we have compared the two years (2009, 2010) of ROMS numerical simulations with current meter data, deployed in the Corsica Channel. Data has been collected by the `Istituto di Scienze Marine Ð Consiglio Nazionale delle Ricerche' (ISMAR-CNR, La Spezia). We considered data from two moorings, located at  $9.685^{\circ}$ E, Longitude and $43.033^{\circ}$ N, Latitude, measuring at two different depths (70m, 125m). From March 2010, the current meters have been replaced by an Acoustic Doppler Current Profiler (ADCP), which measured currents from 13m to about 380m, processed in 20m bins. During the study period, we compared the simulated current in the Corsica Channel with current meters and ADCP data. The model equivalent mean current, are in good agreement with the measured data, for the same approximate location. The current between 70-125m has the same general direction (north), and comparable magnitudes (0.4-0.2 $ms^ {-1}$ in winter; $0.1-0.2 ms^ {-1}$ during summer), as well as the same seasonal variability. ROMS also reproduced the general SST and altimetry, monthly and seasonal trends, measured with satellite sensors.

\subsection{Eddy detection algorithm}
In order to detect and track eddies from our numerical solution, the `Find Okubo-Weiss Eddies in Aviso SSH Product tool contained in the MGET tool for ArcGis' \citep{Roberts_etal_2010_EMS} was adapted to be used with the ROMS output. As defined by \citet{Henson_Thomas_2008_DSR}, an eddy consists of a region of high vorticity (the core), surrounded by a circulation cell of high strain (the ring). Such regions can be detected using the Okubo-Weiss parameter \citep{Okubo_1970_DSR, Weiss_1991_PDNP}. The eddy detection algorithm used the Okubo-Weiss parameter (Q) computed for ROMS with Romstools \citep{Penven_etal_2008}, and filter the candidate eddy cores using various techniques designed to remove small, ephemeral eddies and other unwanted features. These techniques include removing eddies that are smaller than a minimum size and tracking eddies through the time-series, and removing eddies that are too short in duration. The final analysis classifies eddies as either anticyclonic or cyclonic based on the curvature of the sea surface height (anticyclonic is concave down, cyclonic is concave up). After a sensitivity analysis for the study region, a threshold value of Q = ($-1.5e^{-9} s^{-2}$) was chosen, and only eddies with a minimum duration of 3 days, and a minimum area of 8 pixels (about 5km radius of the vorticity core), were considered.

\section{The Productive Ligurian Pool}

\begin{figure}
 \center
\includegraphics[scale=0.45]{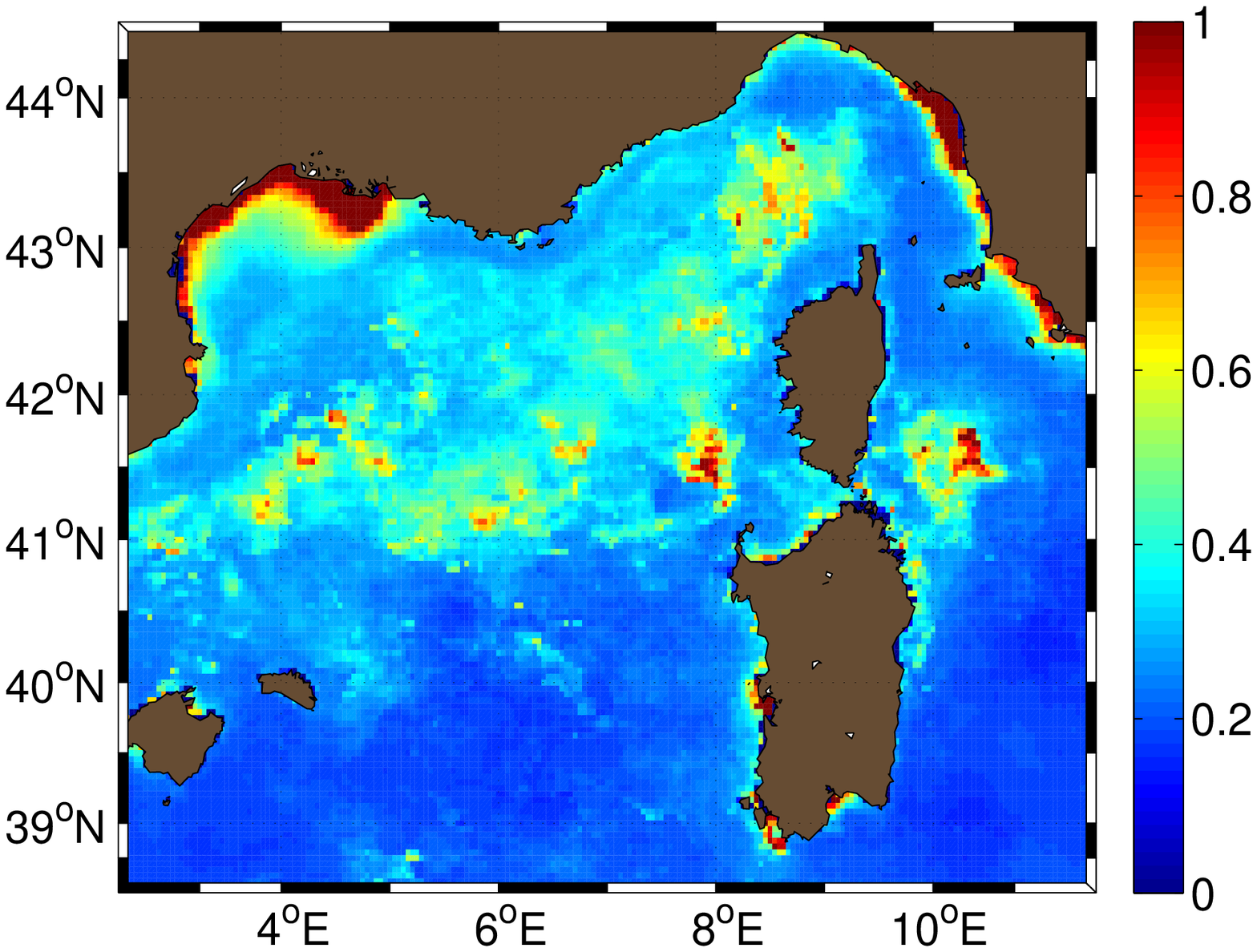}
\includegraphics[scale=0.45]{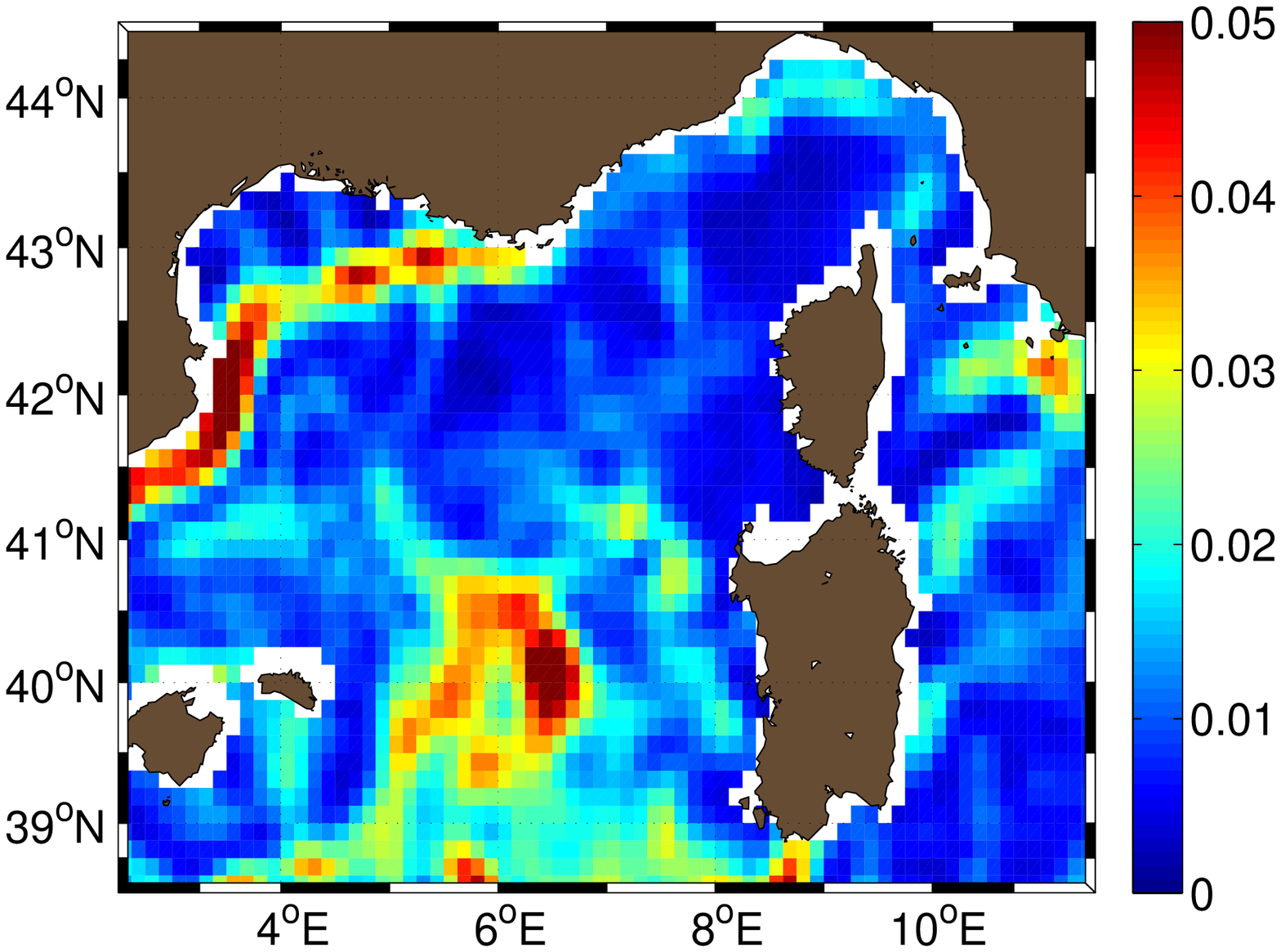}
\includegraphics[scale=0.45]{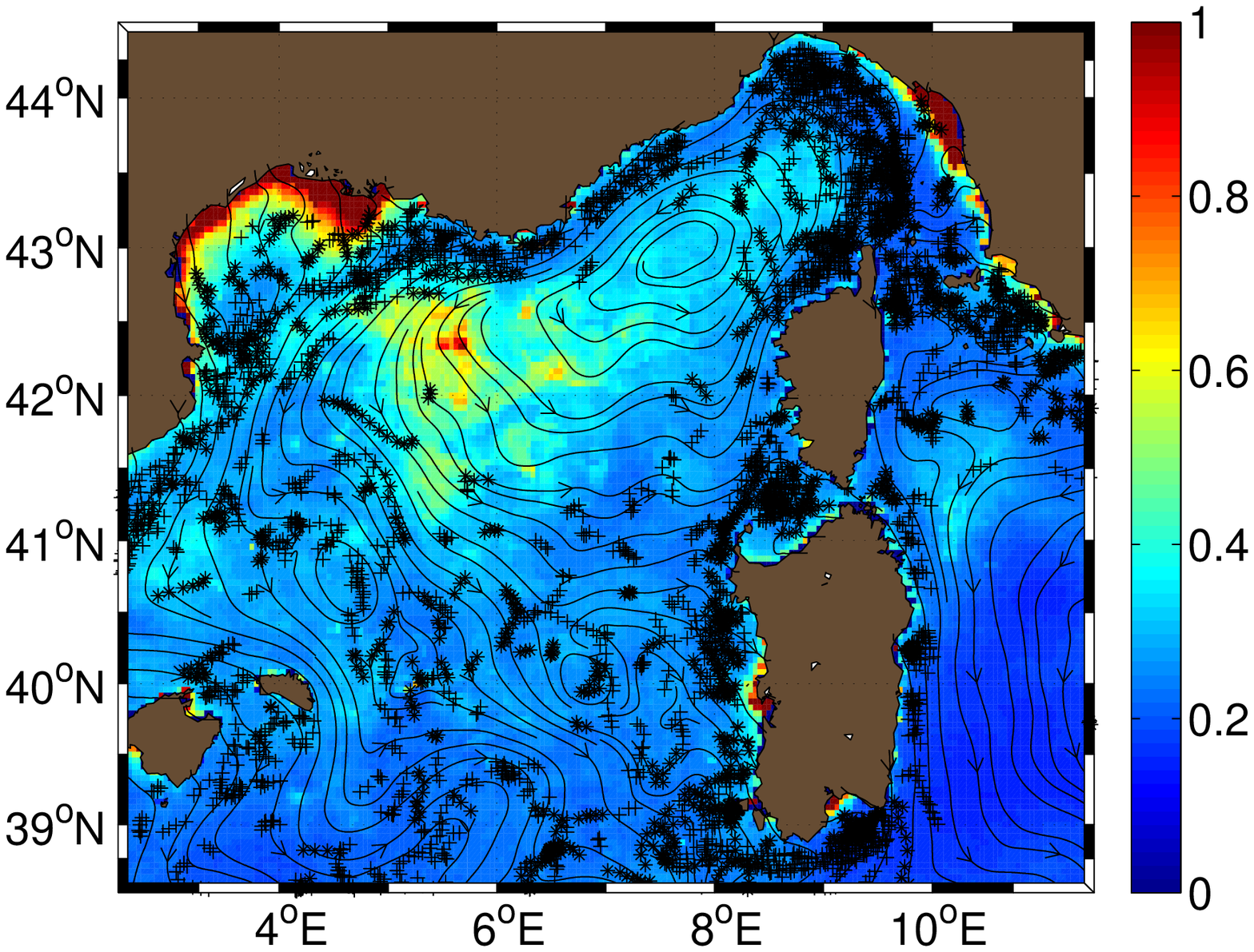}
\includegraphics[scale=0.45]{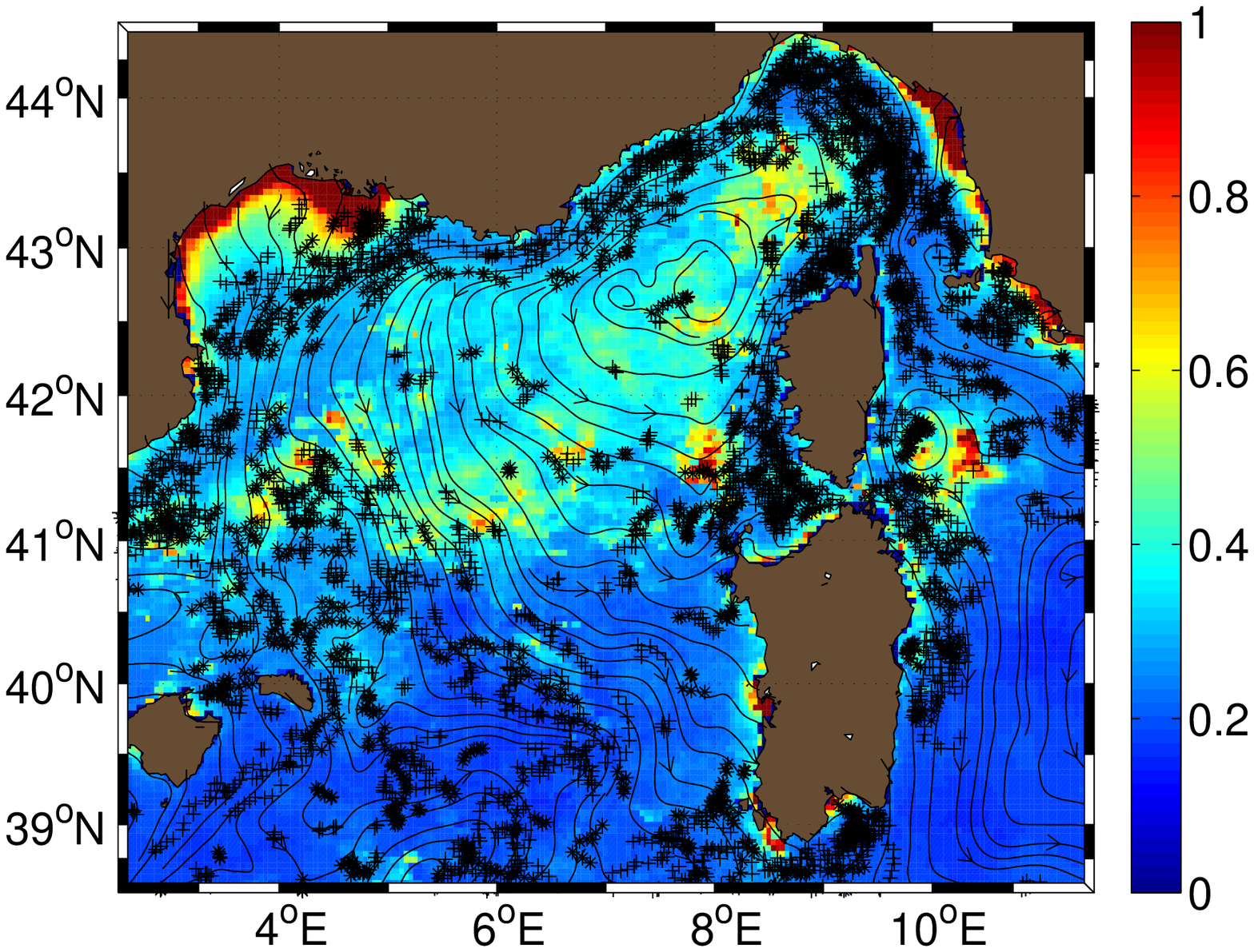}
\caption{Top Panel: (a) Yearly averaged chlorophyll ($mg m^{-3}$), derived from MODIS-Aqua; and (b) yearly averaged EKE ($m^{2} s^{-2}$) derived from AVISO altimetry, for 2010. Bottom panel: Yearly averaged ROMS streamlines and eddy detected (black spots),representative of the averaged circulation for the Ligurian-Proven\c{c}al Basin for (c) 2009 and for (d) 2010. Mode solutions are overlaid onto yearly averaged chlorophyll data}
\label{chleke}
 \end{figure}

The analysis of yearly mean ocean color data, confirmed the persistent existence of productive zone in centre of the Ligurian Sea, during 2009 and 2010. Figure \ref{chleke} shows the spatial distribution of the yearly averaged chlorophyll concentration in  2010, (2009 has a similar pattern, figure \ref{chleke}-c).  The analysis of the concurrent altimetry data for 2010 also suggests that the region surrounding the LPB has the most intense EKE values (Figure \ref{chleke}-b). The yearly averaged chlorophyll concentration maintains the imprint of the spatial distribution during the spring bloom. Figure \ref{chleke} shows a dynamic region at the rim dominated by the ECC, the WCC and the LPC-current, with a low chlorophyll concentration, whereas the central area (Productive Ligurian Pool), shows a high chlorophyll concentration. In fact, the ROMS simulations confirm the presence of the well known cyclonic gyre surrounding the LPB, as shown by streams lines for the 2009 and 2010 model solutions (Figure \ref{chleke}). The model clearly shows a cyclonic circulation which surrounds the productive area, with intense eddy activity in the rim. To note is the inter-annual variability, the main patch of chlorophyll is not located in the same location inside the cyclonic gyre, thus suggesting the heterogeneity previously discussed in the literature.

Eddies are located along the path of the ECC, the WCC and the LPC currents, as it is shown in figure \ref{chleke}. A highly dynamic rim of eddies can permit the transport of nutrients from the eddy-induced vertical mixing and/or advected from a nearby coastal regions to be contained in the central, and less turbulent zone (hence the designative name of pool). This transport is expected to have a strongest  enriching effect when MLD in the central zone is shallower and nutrient can remain in the euphotic layer (i.e. during spring). Thus it is not surprising that productivity is twice as high during the spring bloom, when compared to the yearly averaged values, nevertheless the general pattern is maintained. During 2009, 810 eddies were detected in the realistic ROMS solutions of the LPB, 316 (40\%) were cyclonic and 485 (60\%) anticyclonic; whereas during the 2010 simulation, 935 eddies were counted, 357 (38\%) cyclonic and 578 (62\%) anticyclonic. The strongest wind-stress peaks were also calculated for the 2010 solution, suggesting that the wind plays a role on the generation of (sub)mesoscale features, which in turn seem to condition the overall productivity of the region. In fact, the 2010 spring bloom showed higher chlorophyll concentration values, compared to 2009.

Constricted basins and marginal seas in contrast with the open-ocean can not assume steady-state conditions, hence the recent developments on strait-marginal seas systems theories. \citet{PratSpall_2008_JPO}, used numerical models to test a theory, which explains the exchanged occurring between a schematic buoyancy-forced marginal sea and the open ocean. The incoming surface layer formed a baroclinic unstable boundary current, that circles the marginal sea in a cyclonic manner and feeding heat to the interior, by way of mesoscale eddies. As in the Ligurian-Proven\c{c}al Basin, eddies play a fundamental role in the maintenance and in the exchange processes with the marginal sea. Consistent with the overall heat and volume balances for the marginal sea, \citet{PratSpall_2008_JPO} found a continuous family of hydraulically controlled states including critical flows, mediated by eddies, in a strait between the marginal sea and the open-ocean. Nevertheless, the behavior of eddies in a constricted basin and/or marginal sea is not yet fully understood, further investigation of the LPB could continue to provide new insights. \citet{Kida_etal_2009_JPO} showed that  the eddies accompanying baroclinic instability in the Faroe Bank Channel can set up a double-gyre circulation in the upper ocean, an eddy-driven topographic beta plume. In regions where baroclinic instability is growing, the momentum flux from the overflow into the upper ocean can act as a drag on the overflow causing the overflow to descend the slope at a steeper angle than what would arise from bottom friction alone.  In contrast, the upper layer of the Mediterranean overflow is likely to be dominated by an entrainment-driven topographic beta plume. The difference arises because entrainment occurs at a much shallower location for the Mediterranean case and the background potential vorticity gradient of the upper ocean is expected to be much larger \citep{Kida_etal_2009_JPO}.


%
%
%
%
%
%
%

\begin{acknowledgments}
The authors are grateful for travel funds provided by CIMA Foundation, which enabled researchers to materialized this collaboration, and to CCM for hosting the researchers in Madeira. Numerical model solutions were calculated at CIIMAR HPC unit, constructed using funds the FCT-Portuguese National Science Foundation pluriannual, and from RAIA ($0313\_RAIA\_1\_E$) and RAIA.co projects, co-funded by INTERREGÐIV and by FEDER (`Fundo Europeu de Desenvolvimento Regional, 2007Ð2013'),  through the POCTEP regional initiative. The altimeter products were produced by SSALTO/DUACS and distributed by AVISO with support from CNES. MODIS-Aqua data was extracted using the Giovanni online data system, developed and maintained by the NASA GES DISC.
\end{acknowledgments}

%
%
%
%
%
%
%
%
%

\bibliography{PLPbib}
\bibliographystyle{agu08}






%

%
%

\end{article}




%

%
%
%
%
%
%


\end{document}